\documentclass{iopconfser}

\usepackage{orcidlink}
\usepackage{amsmath}%
\usepackage{amsfonts}%
\usepackage{amssymb}%
\usepackage{graphicx}
\usepackage{xspace}
\usepackage{amsthm}

\newcommand{\orcid}[1]{\href{https://orcid.org/#1}{\textcolor[HTML]{A6CE39}{\aiOrcid}}}

\def\({\big(}
\def\){\big)}
\newcommand{\tn}{\textnormal}

\newcommand{\ket}[1]{|#1\rangle}
\newcommand{\braket}[2]{\langle#1|#2\rangle}
\newcommand{\hilbert}{\mathcal{H}}
\newcommand{\wh}[1]{\widehat{#1}}
\newcommand{\opern}[1]{\mathbf{#1}}
\newcommand{\oper}[1]{\wh{\opern{#1}}}
\newcommand{\R}{\mathbb{R}}

\newcommand{\schrod}{Schr\"odinger\xspace}
\newcommand{\q}{\boldsymbol{q}}

\newcommand{\dof}{d.o.f.\xspace}
\newcommand{\dofs}{d.o.f.s\xspace}

\newcommand{\xThree}{\boldsymbol{x}}
\newcommand{\pThree}{\boldsymbol{p}}

\newcommand{\n}{\mathbf{n}}
\newcommand{\abs}[1]{\left\lvert#1\right\rvert}

\theoremstyle{definition}
\newtheorem{observation}{Observation}
\newtheorem{puzzle}{Puzzle}
\newtheorem{assumption}{Assumption}

\begin{document}

\title{What makes you an observer?}

\author{Ovidiu Cristinel Stoica\ \orcidlink{0000-0002-2765-1562}}

\affil{Dept. of Theoretical Physics, NIPNE---HH, Bucharest, Romania.}

\email{\href{mailto:cristi.stoica@theory.nipne.ro}{cristi.stoica@theory.nipne.ro},  \href{mailto:holotronix@gmail.com}{holotronix@gmail.com}}

\begin{abstract}
Both empirical and theoretical objective science can only access relations, revealing nothing about the intrinsic nature of the entities ``in relation''. We typically refer to these entities as ``matter'', assuming their nature is irrelevant and that relational structures alone explain all phenomena, including consciousness. This would imply that consciousness arises not from matter itself but solely from its structural configurations. Consequently, if two worlds possess isomorphic structures and obey identical dynamical laws, isomorphic systems within these worlds should be equally sentient or insentient.

I demonstrate that if this premise held true, the memories within an observer's brain would bear no correlation to the external world. Yet since we do acquire knowledge of the external world, structure alone must be insufficient: there exists something beyond relations, the intrinsic essence of the ``stuff in relation''. This essence imbues the observer's structure with sentience, metaphorically ``breathing fire'' into it. In turn, the observer confers physical meaning upon the world's structures.

Remarkably, this conclusion emerges directly from physics and mathematics. Without acknowledging this intrinsic element, we would be incapable not only of subjective experience but also of acquiring any objective knowledge about the physical world.
\end{abstract}

\section{Introduction}
\label{s:intro}

We are part of the reality we observe and try to describe. This raises significant challenges. One of them, which is not the central focus of this article, though we will return to it in closing, is that, due to our active involvement in observations, it has proven extremely difficult to explain why superpositions are never directly observed in quantum experiments, even when systems appear prepared in such states. While Bohr attributed this to the classical nature of measuring devices \cite{Bohr1928ComoLecture,Bohr1935CanQuantumMechanicalDescriptionOfPhysicalRealityBeConsideredComplete}, Heisenberg argued that measurements terminate not at the apparatus but at the observer \cite{Heisenberg1958PhysicsAndPhilosophy}, a view later endorsed by von Neumann \cite{vonNeumann1955MathFoundationsQM}, Wigner \cite{Wigner1967RemarksOnTheMindBodyProblemWignersFriend}, and others. Conversely, alternative formulations of quantum mechanics have sought to eliminate observers entirely \cite{Bohm1952SuggestedInterpretationOfQuantumMechanicsInTermsOfHiddenVariables,Everett1957RelativeStateFormulationOfQuantumMechanics,Bell1986QFTWithoutObservers,Goldstein1998QuantumTheoryWithoutObserversI,Goldstein1998QuantumTheoryWithoutObserversII,Stoica2020StandardQuantumMechanicsWithoutObservers}.

This article identifies a distinct observer problem, arising from the assumption that physical reality is exhaustively characterized by structure. It is widely held that reality, including observers and their minds, is fully defined by structural properties, as experiments and mathematical models access only relations (structure), not the intrinsic nature of the relata. While rarely stated explicitly, this view underpins how physicists treat equivalent formulations derived from distinct foundational structures and the assumption that simulating consciousness suffices to produce it. The arguments here apply to approaches grounded, explicitly or implicitly, in this premise.

We demonstrate that this assumption leads to an unexpected conclusion: an observer could not know anything about the external world. Since this contradicts all empirical evidence and everyday perception, the hypothesis that physical meaning derives solely from structure is refuted.

This has profound implications for physics, rejecting research programs that seek to derive physical entities (a preferred pointer state, a preferred basis, subsystem decomposition, or spacetime) from quantum-mechanical structures or relations alone.

Regarding the observers, the result challenges reductionist claims that the mind is reducible to computation \cite{sep-computational-mind} or purely functional properties of material systems \cite{sep-functionalism,sep-physicalism,sep-materialism-eliminative}.

This conclusion may seem surprising, as questions about the nature of consciousness are often dismissed as metaphysical. Yet it follows directly from mathematical physics and empirical observation, requiring no metaphysical assumptions.

\section{The structure of a physical system}
\label{s:system-structure}

Observers are subsystems of the quantum universe, so we begin here. The physical state of the quantum universe is represented by a state vector $\ket{\psi(t)}$, a unit vector in a complex Hilbert space endowed with a Hermitian scalar product. Its dynamics are governed by the \schrod equation:
 \begin{equation}
	\label{eq:unitary_evolution}
	\ket{\psi(t)}=e^{-\frac{i}{\hbar}(t-t_0)\oper{H}}\ket{\psi(t_0)},
\end{equation}
where $\oper{H}$ is the \emph{Hamiltonian operator}. Since the operator $\oper{H}$ is Hermitian, $e^{-\frac{i}{\hbar}(t-t_0)\oper{H}}$ is unitary.

But a unit vector is just like any other unit vector, so how are the extremely diverse possible physical states supposed to be represented by a unit vector? We can distinguish them by their different physical properties. A physical property $\opern{A}$ is represented by a Hermitian operator $\oper{A}$. A system in a state $\ket{\psi(t)}$ has a definite value $a\in\R$ for some physical property $\opern{A}$ if $\ket{\psi(t)}$ is an eigenvector of the operator $\oper{A}$ with the eigenvalue $a$, that is, $\oper{A}\ket{\psi(t)}=a\ket{\psi(t)}$.

Subsystems of the universe can be characterized by their physical properties. If the two subsystems don't have common subsystems, the operators representing the properties of a system $S_1$ commute with those of another system $S_2$. The universe can thus be decomposed into subsystems, and the state space of the universe is represented as the tensor product of the state spaces of the subsystems. We call this decomposition the \emph{tensor product structure}. The algebras of observables of the subsystems fully determine the tensor product structure of a system consisting of $n$ elementary particles
 \cite{ZanardiLidarLloyd2004QuantumTensorProductStructuresAreObservableInduced}.

Consider the $j$-th elementary particle from a non-relativistic system of $n$ elementary particles. This particle is characterized by some degrees of freedom (\dofs), including space coordinates $(x_j,y_j,z_j)$, and spin and internal \dofs $(w_{j1},\ldots,w_{jd_j})$. 
These \dofs form a configuration space whose points are
\begin{equation}
\label{eq:config-n}
\q:=\(q_1,q_2,\ldots\)=\(\underbrace{x_1,y_1,z_1,w_{11},\ldots,w_{1d_1}}_{\tn{particle }1},\ldots,\underbrace{x_n,y_n,z_n,w_{n1},\ldots,w_{nd_n}}_{\tn{particle }n}\).
\end{equation}

These \dofs are the possible values of some physical properties, or the possible eigenvalues of the operators representing them.
The common eigenvectors of these operators form a basis of the state space of $n$ particles. These basis vectors can be labeled by the eigenvalues, which form the coordinates in the configuration space,
\begin{equation}
\label{eq:basis-n}
\ket{\q}=\ket{q_1,q_2,\ldots}=\ket{\underbrace{x_1,y_1,z_1,w_{11},\ldots,w_{1d_1}}_{\tn{particle }1},\ldots,\underbrace{x_n,y_n,z_n,w_{n1},\ldots,w_{nd_n}}_{\tn{particle }n}}.
\end{equation}

The wavefunction of a system of $n$ particles consists of the components of the state vector
\begin{equation}
\label{eq:wf-n}
\psi(\q,t)=\braket{\q,\ldots}{\psi(t)}\text{ or }\psi(q_1,q_2,\ldots,t)=\braket{q_1,q_2,\ldots}{\psi(t)}
\end{equation}
in the basis \eqref{eq:basis-n}.
It propagates on the configuration space of the $n$ particles.

This description works for both non-relativistic quantum mechanics and quantum field theory formulated in the Fock space (if we sum over states of different numbers of particles). Equivalently, there is a wavefunctional formulation of quantum field theory in which $q_j$ represent classical field configurations rather than the \dofs of each particle \cite{Hatfield2018QuantumFieldTheoryOfPointParticlesAndStrings}. A similar formulation can be used for quantum gravity \cite{Dewitt1967QuantumTheoryOfGravityI_TheCanonicalTheory}.

Each classical position operator $\hat{q}_j$ has a canonical conjugate operator, the momentum operator
\begin{equation}
\label{eq:momentum_j}
\hat{p}_j:=-i\hbar\frac{\partial}{\partial q_j}.
\end{equation}

If $q_j$ represents a classical field configuration, the canonical conjugate of the field operator $\hat{q}_j$ is the operator given by the functional derivative $\hat{p}_j:=-i\hbar\frac{\delta}{\delta q_j}$.

All other operators are functions of the position (or field) operators and their canonical conjugates, as well as the spin and internal \dof operators.

If quantum mechanics is complete, everything there is to know about the world, in particular every observer, is encoded in the wavefunction, that is, in the state vector and the operators $\(\hat{q}_1,\hat{q}_2\ldots\)$, \emph{if} we know what physical properties they represent.

\begin{observation}
\label{obs-schrod-daemon}
A hypothetical superintelligent being (let's call such a being the \emph{\schrod daemon} for the similarity with the Laplace daemon) should then be able to read everything there is to know about the world from the patterns of the wavefunction. In particular, it should be able to recognize subsystems and their physical properties, including observers with their thoughts and sentient experiences.
\end{observation}

\section{Do we need more than structure?}
\label{s:structuralism}

We gather knowledge about the world through measurements and observations. When measuring a quantity, we compare it to conventional standards: distances against inches or feet, masses against reference objects, and so on. Ultimately, measurements yield ratios, relations between quantities. Other observations involve counting discrete entities, such as particles produced in collisions. Every measured property reduces to numerical values or identifiable patterns. In all cases, we access relations, not intrinsic natures.

When systematizing knowledge into mathematical models, we again encounter relations. A theory's most mature form is a mathematical structure (sets with relations between them \cite{Gratzer2008UniversalAlgebra,ChangAndKeisler1990ModelTheory}). This led philosophers to conclude that scientific knowledge encompasses only structure (relations between theory elements), not the relata's intrinsic nature \cite{Worrall1989StructuralRealism,sep-structural-realism,Wallace2022StatingStructuralRealismMathematicsFirstApproachesToPhysicsAndMetaphysics}. The latter, we think, remains metaphysical.

Though physicists may intuit internal representations of physical objects, they avoid referencing intrinsic natures, they try to come up with mathematical
models, which are inherently relational. When two distinct models admit a one-to-one correspondence so that all statements about one of them are translatable to the other, they are deemed equivalent formulations of the same theory. Thus, classical mechanics is equally describable via Newtonian, Lagrangian, or Hamiltonian frameworks. Quantum mechanics similarly permits multiple formulations: Heisenberg, Schr\"odinger, and Dirac pictures; field operators, wavefunctionals, path integrals, and Feynman diagrams. Likewise, gauge-gravity duality posits physical equivalence between a bulk spacetime theory and a lower-dimensional boundary theory \cite{Maldacena1999AdSCFTInitial,Wallace2022StatingStructuralRealismMathematicsFirstApproachesToPhysicsAndMetaphysics}.

These considerations motivate:
\begin{assumption}
\label{assumption:structuralism}
All there is to know about the physical world is encoded in structure.
\end{assumption}

Ignoring intrinsic natures has proven productive. Consider space: we describe relative positions and coordinate-dependent changes but remain silent on space's intrinsic nature or point identities over time. This relational focus enabled discoveries like special/general relativity and gauge theories. Unsurprisingly, physicists adopt, explicitly or implicitly, a relational worldview.

\section{Puzzle: the ambiguity of the physical meaning}
\label{s:ambiguity-structure}

The logical conclusion of relationism, that only structure matters, has led many physicists to propose that a complete description of reality requires only the state vector (the unit vector, not the wavefunction, which is the state vector's components in a particular basis associated with a physical meaning) and the Hamiltonian (in its abstract form, that is, its spectrum alone). Carroll and Singh termed this view \emph{Hilbert-space fundamentalism} \cite{CarrollSingh2019MadDogEverettianism}, p. 95:
\begin{quote}
Typically one also relies on some classical structure, such as space and local configuration variables within it, which then gets promoted to an algebra of preferred observables. We argue that even such an algebra is unnecessary, and the most basic description of the world is given by the spectrum of the Hamiltonian (a list of energy eigenvalues) and the components of some particular vector in Hilbert space. Everything else---including space and fields propagating on it---is emergent from these minimal elements.
\end{quote}

This principle, implicitly adopted before and after Carroll and Singh's work, underpins claims that decoherence dynamically selects a preferred basis \cite{Zurek1991DecoherenceAndTheTransitionFromQuantumToClassical,Zurek2003DecoherenceEinselectionAndTheQuantumOriginsOfTheClassical,Schlosshauer2007DecoherenceAndTheQuantumToClassicalTransition}, that, under locality constraints, the Hamiltonian uniquely determine tensor product structures \cite{Piazza2010GlimmersOfAPreGeometricPerspective,CotlerEtAl2019LocalityFromSpectrum,CarrollSingh2021QuantumMereology}, and that space itself uniquely emerges from abstract quantum structures \cite{CarrollSingh2019MadDogEverettianism,Carroll2021RealityAsAVectorInHilbertSpace,CarrollSingh2021QuantumMereology}.

However, none of these structures emerge uniquely \cite{Stoica2022SpaceThePreferredBasisCannotUniquelyEmergeFromTheQuantumStructure,Stoica2023NoGoResultsOnEmergentSpaceAndOtherStructures,Stoica2024EmpiricalAdequacyOfTimeOperatorCC2HamiltonianGeneratingTranslations,Stoica2023PrinceAndPauperQuantumParadoxHilbertSpaceFundamentalism,Stoica2024DoesTheHamiltonianDetermineTheTPSAndThe3dSpace}. The general proof in \cite{Stoica2022SpaceThePreferredBasisCannotUniquelyEmergeFromTheQuantumStructure} applies to any quantum structure with any Hamiltonian and any invariant constraints (\emph{e.g.} the constraint of locality). To understand this result we can examine the simpler case of a non-relativistic Hamiltonian 
\begin{equation}
\label{eq:schrod_hamiltonian_NRQM}
\oper{H} = \sum_{j=1}^\n\frac{1}{2m_j}\hat{\pThree}_{j}^2
+ \mathop{\sum_{j=1}^\n\sum_{k=1}^\n}_{j\neq k}\hat{V}_{j,k}\(\abs{\hat{\xThree}_j-\hat{\xThree}_k}\),
\end{equation}
where 
$m_j$ is the mass of the particle $j$,
$\hat{\xThree}_{j}=\(\hat{x}_{j1},\hat{x}_{j2},\hat{x}_{j3}\)$ are its position operators, $\hat{\pThree}_{j}=\(\hat{p}_{j1},\hat{p}_{j2},\hat{p}_{j3}\)$ its momentum operators, and $\hat{V}_{j,k}$ is a potential that depends only on the distances between two positions.

The question of whether a unique space emerges can be stated in the following way: given the Hamiltonian $\oper{H}$ from equation \eqref{eq:schrod_hamiltonian_NRQM}, are the operators $\hat{\xThree}_{j}$ and $\hat{\pThree}_{j}$ the only ones (up to isometries of space and permutations of identical particles) for which $\oper{H}$ has the same form as in equation \eqref{eq:schrod_hamiltonian_NRQM}?

The answer is no! Any operator $\oper{S}$ that commutes with $\oper{H}$ gives us such a set of operators, $\hat{\xThree}'_{j}:=\(\oper{S}\hat{x}_{j1}\oper{S}^\dagger,\oper{S}\hat{x}_{j2}\oper{S}^\dagger,\oper{S}\hat{x}_{j3}\oper{S}^\dagger\)$ and $\hat{\pThree}'_{j}:=\(\oper{S}\hat{p}_{j1}\oper{S}^\dagger,\oper{S}\hat{p}_{j2}\oper{S}^\dagger,\oper{S}\hat{p}_{j3}\oper{S}^\dagger\)$, so that equation \eqref{eq:schrod_hamiltonian_NRQM} remains true when we replace $\oper{H}=\oper{S}\oper{H}\oper{S}^\dagger$, $\hat{\xThree}_{j}$ with $\hat{\xThree}'_{j}$, and $\hat{\pThree}_{j}$ with $\hat{\pThree}'_{j}$.
There are infinitely many such operators $\oper{S}$, in fact infinitely many infinite families of such operators.
And, as shown by the examples provided in \cite{Stoica2022SpaceThePreferredBasisCannotUniquelyEmergeFromTheQuantumStructure,Stoica2023PrinceAndPauperQuantumParadoxHilbertSpaceFundamentalism,Stoica2024DoesTheHamiltonianDetermineTheTPSAndThe3dSpace}, we obtain infinitely many possible solutions that are completely distinct from a physical point of view, and they correspond to physically distinct structures like space and decomposition into subsystems.

The same applies even if our quantum theory is formulated in terms of other fundamental observables than position operators, as shown in full generality in \cite{Stoica2022SpaceThePreferredBasisCannotUniquelyEmergeFromTheQuantumStructure,Stoica2023NoGoResultsOnEmergentSpaceAndOtherStructures}. Any other operators that we may think play a special role in our theory, including in quantum gravity, will have to face the same problems.

Consequently, in \cite{Stoica2022SpaceThePreferredBasisCannotUniquelyEmergeFromTheQuantumStructure,Stoica2023NoGoResultsOnEmergentSpaceAndOtherStructures} I argued that we need to fix an algebra of preferred observables, otherwise the same state vector supports infinitely many physically distinct interpretations as a physical world.

This leads to another question: would the existence of a preferred set of operators solve the problem and give us a complete description of reality from structures alone?
Let's see. First, such operators can specify how the universe is divided into subsystems. Second, by specifying what operators for each subsystem corresponds to space (in the above example these are $\hat{\xThree}_{j}$), it can give us the space structure.

But along with the operators $\hat{\xThree}_{j}$, their conjugate momenta $\hat{\pThree}_{j}$ also exist, because they are partial derivative operators with respect to the eigenvalues of $\hat{\xThree}_{j}$. And with these, all other operators exist, since they are functions of $\hat{\xThree}_{j}$ and $\hat{\pThree}_{j}$. Just like along with two apples and other three apples, five apples exist. So by merely denoting some operators by $\hat{\xThree}_{j}$ and $\hat{\pThree}_{j}$ we don't acquire much, since this is a mere notation by which we only highlight them in our minds among other possible sets of operators that satisfy the same relations.
Their privileged status is not encoded in the relations, since infinitely many other choices satisfy the same relations.
Therefore, we arrived at a puzzle.
\begin{puzzle}
\label{puzzle:ambiguity}
Structure (relations and dynamics) insufficiently determines physical reality: there is no unique way to recover physical properties from operators.
\end{puzzle}

To try to resolve this ambiguity, we turn to observations: after all, first we gather empirical data through observations, and then we systematize them into theories using mathematical structures.

\section{Puzzle: the observer's zero knowledge}
\label{s:zero-knowledge}

If observations can solve the ambiguity of assignment of physical meaning to the operators, we need to include them, along of course with the observers doing them, in our description of the world.

One might assume including the measuring devices suffices. As subsystems of the universe, their full description lies in their wavefunctions (or reduced density matrices)---the state vector's components in eigenbases of position or other physical-property operators, as noticed in Observation \ref{obs-schrod-daemon}. However, this leads to circularity: determining what physical property each operator pertaining to the measuring device represents requires this very operator-physical property correspondence to already exist.

Realizing that the results displayed by the measuring devices eventually reach the observers, who interpret these results, maybe we can break the circularity by including the observers in the description of the world. In fact, we should do this anyway, because even observers who never did a quantum measurement perceive the world through their sense organs, which are also measuring devices.
Yet if observers themselves are subsystems, how is circularity avoided?

Each choice of operators $\hat{\xThree}'_{j}$ to represent positions results in distinct configuration spaces, different eigenbases, and therefore different wavefunctions.
In fact, since for any two possible state vectors $\ket{\psi}$ and $\ket{\psi'}$ there is a transformation $\oper{S}$ so that $\ket{\psi'}=\oper{S}\ket{\psi}$, the wavefunction can look like any other wavefunction, depending on the assignment of physical meaning to the operators! While most such choices alter the Hamiltonian's form \eqref{eq:schrod_hamiltonian_NRQM}, restricting to and preserving \eqref{eq:schrod_hamiltonian_NRQM} still yields vast underdetermination.
This freedom allows infinitely many physically distinct interpretations of the external world.

To see this, recall that an observer is a subsystem of the universe. We notice that the operators $\hat{\xThree}_{j}$ and $\hat{\pThree}_{j}$ are, for most values of $j$, position-like and momentum-like operators for particles that are external to the observer. Assuming as given some operators $\hat{\xThree}_{j}$ and $\hat{\pThree}_{j}$ presumed to represent positions and momenta, there seems therefore to be plenty of freedom to choose an operator $\oper{S}$ that commutes with $\oper{H}$ and leaves the position and momentum operators $\hat{\xThree}_{j}$ and $\hat{\pThree}_{j}$ of the particles constituting the observer unchanged, but changes the operators $\hat{\xThree}_{j}$ and $\hat{\pThree}_{j}$ of the particles from the observer's external world.
In addition to varying the choice of the position and momentum operators for the observer's external particles, there are choices that make the observer itself appear as an isomorphic system as seen in terms of other choices for the position and momentum operators. The observer's wavefunction would have the same patterns, but seen from different bases, or propagating on different choices of the configuration space. I will call \emph{observer-like structure} such a structure having the same structure as an observer.

In any case, Lemma 1 in \cite{Stoica2023PrinceAndPauperQuantumParadoxHilbertSpaceFundamentalism} ensures and exacerbates this underdetermination, by showing that symmetry transformations $\oper{S}$ map measurement outcomes to alternative results.
That is, for a quantum measurement, there is a symmetry transformation $\oper{S}$ mapping the operators making the wavefunction describe that a result was obtained into operators making it describe that another result was obtained.
This implies that, even if we were to fix the operator-physical property correspondence, there are many physically distinct ways to assign the eigenvalues of the operators to the values of the physical properties.

Already the results from \cite{Stoica2023PrinceAndPauperQuantumParadoxHilbertSpaceFundamentalism} imply that the structure alone is insufficient, being compatible with opposite assignments of physical meaning to the operators. But it can be applied to prove even more.

Consider an observer and a physical property of an external object, which the observer thinks they know. For example, the observer may think that in the kitchen there is a red apple on the table. The observer can decide to walk in the kitchen and check that indeed there is a red apple on the table. This scenario allows us to apply the above-mentioned lemma and conclude that if the observer goes to the kitchen, she can very well find a green or a pink apple on the table. It would be all the same for the state vector and the operators, the difference residing only in the operator-physical property correspondence.

The ambiguity extends to shape, location, and even the apple's existence, as particle arrangements admit multiple interpretations.

This ignorance differs fundamentally from classical uncertainty. It doesn't arise not from the usual absence of knowledge of physical details, as when the observer has a red and a pink apple, and she tells a friend on the way out to take one of them, and she doesn't know which apple remained.
It arises purely from structural underdetermination.
And this underdetermination is maximal, because unitary symmetry and uniform measures over operator spaces render all property values equally probable, leaving observers knowing nothing about the external world! For more details, please consult \cite{Stoica2023AreObserversReducibleToStructures}.

This result leads to another puzzle:
\begin{puzzle}
\label{puzzle:zero-knowledge}
Observations don't resolve the ambiguity. Moreover, this ambiguity makes it so that the observers know nothing about the external world.
\end{puzzle}

\section{Attempted solutions of the puzzles}
\label{s:attempts}

Puzzles \ref{puzzle:ambiguity} and \ref{puzzle:zero-knowledge} stem from Assumption \ref{assumption:structuralism}---the claim that physical reality is fully determined by structure. This assumption rests on the empirical and theoretical focus on relations. Nevertheless, since this assumption appeared so far to be sufficient and proved to be very seminal by leading us to discover relativity and gauge theory, resolving these puzzles without abandoning it remains desirable.

The most straightforward solution seems to be to try to make it clear that some of our structures have a more fundamental role than others or that they are more privileged somehow. But they all coexist simultaneously, none of them could come earlier. Together with the operators $\hat{\xThree}_{j}$, the operators $\hat{\pThree}_{j}$ exist as well, and along with them, all other operators, because they all can be expressed as functions of these, $\wh{A}=A\(\hat{\xThree}_{j},\hat{\pThree}_{j}\)$.
It makes no sense to think that somehow $\hat{\xThree}_{j}$ appeared first, then $\hat{\pThree}_{j}$ via \eqref{eq:momentum_j}, then all other operators $\wh{A}=A\(\hat{\xThree}_{j},\hat{\pThree}_{j}\)$.
Declaring some operators ``more fundamental'' is as arbitrary as privileging the numbers $2$ and $3$ over $5$ or $6$.
But maybe the choice of the position and momentum operators is restricted by some special requirement, which is still relational or structural. But since any such condition should be expressed in invariant terms, otherwise it is not basis-independent, it is an arbitrary choice that doesn't actually resolve the problem, it rather smuggles in our description a preferred choice. If it is not invariant, unitary transformations commuting with $\oper{H}$ generate equally valid alternatives \cite{Stoica2022SpaceThePreferredBasisCannotUniquelyEmergeFromTheQuantumStructure}.

Another attempt may come from the observation that, since most observer-like structures isomorphic with our observer have completely wrong knowledge about their environment, these observer-like structures are akin to Boltzmann brains.
A Boltzmann brain is a structure able to hold information like a brain, but which is ephemeral, because it would appear due to an extremely improbable fluctuation \cite{Eddington1934NewPathwaysInScience_Boltzmann_brains}.
The Second Law of Thermodynamics, by requiring that the universe started in a very special state, should reduce the probability of the appearance of a Boltzmann brain.
Since most observer-like structures isomorphic with a real observer would be akin to Boltzmann brains, one may hope that the Second Law of Thermodynamics can help us reduce their occurrence so that the probability that you are an observer-like structure with the wrong knowledge about the exterior world is significantly reduced.
But this is not the case, because we already set our argument in our universe, which is already subject to the Second Law of Thermodynamics, and yet we obtained the existence of infinitely many such observer-like structures that know mostly the wrong properties of the world.
For every observer with the right knowledge, there are infinitely many observer-like structures, in other bases, whose brains store the wrong values for the properties of the external world.
Most of them are, in their bases, like Boltzmann brains.
Therefore, while Puzzle \ref{puzzle:zero-knowledge} may remind us of Boltzmann brains, observer-like structures with the wrong knowledge are unavoidable and significantly dominate those with the right knowledge, regardless of the initial conditions of the universe.
For another way to understand this, suppose that the initial state vector $\ket{\psi_0}$ of the universe was constrained to a subspace $\hilbert_0$ (of states of very low entropy) of the total state space $\hilbert$, and that in the position basis the corresponding wavefunction is very special. Any other possible initial state vector $\ket{\psi_0'}\in\hilbert_0^\perp$, where $\hilbert_0\oplus\hilbert_0^\perp=\hilbert$ has a non-special wavefunction, in the sense that its time evolution results in violations of the Second Law of Thermodynamics. Since $\hilbert_0$ is a subspace of $\hilbert$, there are unitary transformations $\oper{S}$ commuting with $\oper{H}$ so that, in the resulting position-like basis, the wavefunction of $\ket{\psi_0}$ looks totally ``non-special'', that is, it looks like another wavefunction corresponding to a vector from $\hilbert_0^\perp$ expressed in the position basis.
Whatever constraints the wavefunction $\braket{\q}{\psi_0}$ must satisfy, this doesn't eliminate the other bases in which $\braket{\q'}{\psi_0}$ doesn't satisfy the same constraints.
Therefore, appealing to the fact that the initial state of the universe had to be very special to ensure the Second Law doesn't avoid Puzzles \ref{puzzle:ambiguity} and \ref{puzzle:zero-knowledge}.

Similarly, we can try other ways to explain why an observer-like structure will know their environment, for example that natural selection or any other natural cause led to this. But as long as such arguments rely on structure, there will always be unitary transformations resulting in infinitely many observer-like structures isomorphic with an observer but whose brain-structures don't hold the right information about the environment. We can invoke the fact that, in this case, the observer-like structure will soon be destroyed, but at that time there will be other infinitely many observer-like structures isomorphic with the observer but containing the wrong information about the environment.

Ultimately, Puzzles \ref{puzzle:ambiguity} and \ref{puzzle:zero-knowledge} defy resolution under Assumption \ref{assumption:structuralism}, because no constraints prevent the observer-like structures with wrong data in their brain-like structures to exist.

\section{Solution: grounding the physical properties in the observer}
\label{s:grounding}

We've seen that we must abandon Assumption \ref{assumption:structuralism}, because it leads to ambiguities in operator-physical property correspondence, culminating in observers knowing nothing about the external world.

This necessitates positing something beyond structure: the intrinsic nature of relata must break unitary symmetry in state space. This may seem unsettling for science, which is rooted empirically and theoretically in relations. But we make all the time observations that falsify structuralism, because they show that we know about the external world. Scientists and philosophers may deem intrinsic natures ``unscientific'', assuming that they can't be tested, but we've seen that their necessity follows from physics.
We just tested Assumption \ref{assumption:structuralism} and found that there is no way to keep it.

But how would abandoning Assumption \ref{assumption:structuralism} solve the problem? We can't simply state that the position operators $\hat{\xThree}_{j}$ are physically different from any of the operators $\hat{\xThree}'_{j} :=\oper{S}\hat{\xThree}_j\oper{S}^\dagger$ obtained by transforming with $\oper{S}$, because the observer-like structures in any of the bases obtained from eigenvectors of the operators $\hat{\xThree}'_{j}$ would consider the $\hat{\xThree}'_{j}$-basis to be the real position basis! 
Trying to solve Puzzle \ref{puzzle:ambiguity} directly didn't work, but it led us to Puzzle \ref{puzzle:zero-knowledge}, that most observer-like structures would have zero information about the environment. This Puzzle shows that the problem not only persists, but it is worse than we thought. 

Since most observer-like structures have zero information about the environment, and since the reader does know with certainty many properties of the environment, it follows that something prevents the reader from being just any such observer-like structure. Only those observer-like structures that appear as such in a particular position basis can be sentient observers!

At first sight, it may be tempting to say ``but this is obvious, we always knew it!''. But consider the following scenario. Suppose our technology reaches a level that allows us to simulate a human being. Would this simulation be sentient?
At least the supporters of the computational theory of mind would say ``yes!'' \cite{sep-computational-mind,Piccinini2004FunctionalismComputationalismAndMentalStates}. Others would think that a computation is not capturing the essence of being sentient, but if we could capture in our simulation the causal relations underlying our mental processes, the functionality, the simulation would be sentient, as in the functionalist theories of mind \cite{sep-functionalism}.
In any case, if a classical simulation of a person can't be sentient even in principle, it is tempting to think that a quantum simulation should be, at least if it implements as physical cause and effect relations the cause and effect relations in the brain of that person.
If necessary, we can consider simulating the person down to the level of particles, this should be enough, one may think.
But this is exactly what the observer-like structures are, they are simulations down to any discernible structure or relation making up an observer.
And yet, to solve Puzzle \ref{puzzle:zero-knowledge}, we have to refute the possibility that all of them can be sentient observers, in fact, we have to accept that only a zero-measure subset of the set of all possible observer-like structures can be sentient!
Only a minority of them can be sentient, even if they are isomorphic with sentient observers.
This mirrors philosophical zombie arguments, according to which it is conceivable a world containing systems structurally and behaviorally identical to us, yet insentient.
Except that we achieved both the sentient and the insentient structures in the same world, albeit in different bases.

This opens a can of other puzzles, perhaps greater ones. But we know we are sentient, we know we know about the external world, we know what the math tells us, and therefore we have to accept this conclusion. There is something beyond structure, beyond relations, intrinsic to the relata, that gives physical meaning to the operators and without which we would be ``empty'', insentient structures.

This contradicts the view of some physicalists. Sean Carroll wrote in 2021 \cite{Carroll2021ConsciousnessAndTheLawsOfPhysics},
\begin{quote}
By ``physicalism'' here I don't simply mean a world with only physical
properties, but specifically physicalism about consciousness. Can we conceive of a world where the ontology consists of nothing other than some notion of physical ``stuff'' (or the specific quantum fields in the Core Theory) without any inherently mental aspects, but which nevertheless accounts for consciousness as we experience it?
\end{quote}

In 1921, $100$ years before Carroll's account of physicalism, Eddington wrote \cite{Eddington1921SpaceTimeAndGravitation}:
\begin{quote}
And yet, in regard to the nature of things, this knowledge is only an empty shell---a form of symbols. It is knowledge of structural form, not knowledge of content. All through the physical world runs that unknown content, which must surely be the stuff of our consciousness.
\end{quote}

We can, if we want, adopt a more sentience-friendly notion of physicalism, in which the stuff that makes the physical world matters precisely for sentience.
According to Strawson \cite{Strawson2006RealisticMonismWhyPhysicalismEntailsPanpsychism}, page 3:
\begin{quote}
You're certainly not a realistic physicalist, you're not a real physicalist, if you deny the existence of the phenomenon whose existence is more certain than the existence of anything else: experience, `consciousness', conscious experience, `phenomenology', experiential `what-it's-likeness', feeling, sensation, explicit conscious thought as we have it and know it at almost every waking moment.
\end{quote}

Crucially, in this article we derived sentience's necessity from physics, not from introspection, feelings, qualia, `what-it's-likeness'. Without intrinsic content, structural ambiguity precludes both the operator-physical property correspondence and the observer's knowledge about the external world. If I'd define sentience I'd say that it is what makes an observer-like structure be an observer, and what makes an operator correspond to a particular physical property.

\section{Bonus: restoring unitary evolution without many-worlds}
\label{s:collapse}

While focused on a distinct observer problem, our results impact the other observer problem that appears in the measurement problem. Measurements seemingly disrupt unitary evolution \eqref{eq:unitary_evolution}, implying collapse \cite{vonNeumann1955MathFoundationsQM,GhirardiRiminiWeber1980AGeneralArgumentAgainstSuperluminalTransmissionThroughQuantumMechanicalMeasurementProcess}. Or, if the many-worlds interpretation \cite{Everett1957RelativeStateFormulationOfQuantumMechanics} or the de Broglie-Bohm theory \cite{Bohm1952SuggestedInterpretationOfQuantumMechanicsInTermsOfHiddenVariables} is correct, the wavefunction just decoheres.
However, some authors didn't lose the hope that the wavefunction is real and evolves unitarily, so that even when measurements take place, it doesn't collapse, and also doesn't branch, This is the case of Schulman's ``special states'' \cite{Schulman1984DefiniteMeasurementsAndDeterministicQuantumEvolution,Schulman2017ProgramSpecialState}, 't Hooft's cellular automata with discrete time \cite{tHooft2016CellularAutomatonInterpretationQM,Elze2024CellularAutomatonOntologyBitsQubitsAndTheDiracEquation}, and Stoica's ``delayed initial conditions'' and ``global consistency condition'' \cite{Stoica2008SmoothQuantumMechanics,Stoica2021PostDeterminedBlockUniverse}.
Part of the resistance against such proposals is that, to obtain a definite outcome instead of a superposition by unitary dynamics alone, the initial conditions of the universe have to be extremely fine-tuned \cite{Stoica2012QMQuantumMeasurementAndInitialConditions}. This is what Bell called ``superdeterminism'', which can be interpreted alternatively as retrocausality.

It was noticed previously that branching or the need to invoke collapse can be avoided by allowing the initial conditions to be incompletely specified at the Big Bang, and then gradually refined with each new measurement \cite{Stoica2008SmoothQuantumMechanics,Stoica2012QMGlobalAndLocalAspectsOfCausalityInQuantumMechanics,Stoica2016OnTheWavefunctionCollapse,Stoica2017TheUniverseRemembersNoWavefunctionCollapse,Stoica2021PostDeterminedBlockUniverse}.
But the result discussed in this article offers another possibility, that the abstract unit vector is completely specified all the time, but the operator-physical property correspondence is incompletely specified at the Big Bang, and gradually realized with each new observation \cite{Stoica2024ObservationAsPhysication}. 
If we assume that the operator-physical property correspondence existed since the Big Bang, we run into the superdeterminism/retrocausality issues, and it appears that either the observers' decisions are predetermined, or that observers can act retrocausally. But if this correspondence is achieved gradually, with each new observation, extending as it uncovers the world from known to unknown, collapse and branching are avoided.

This also suggests a solution to another problem. Due to the apparent impossibility to intervene in the causal chain, even defenders of consciousness as fundamental tend to give it a purely epiphenomenal place.
This is in stark contrast with our sense that we are sentient and play an active role in the world, because if consciousness is epiphenomenal, how could it be the reason why we can emit mechanical waves or type characters in which we declare that we are sentient?
How could we have free will?
But our observations, we see now, play a double role: on one hand, they affect the world, and on the other hand, they ground the operator-physical property correspondence, while allowing a freedom to select the outcome of an observation.
Therefore, there is room for libertarian free-will despite the determinism of equation \eqref{eq:unitary_evolution} \cite{Stoica2008FlowingWithAFrozenRiver,Stoica2024FreedomInTheManyWorldsInterpretation}.
There is no need for consciousness to break the causal chain in order to act in the physical world, since it can obtain the same results by ``coloring'' the preexisting cold structure given by the unit vector with the physical meaning resulting from gradually grounding the operators in physical properties.



\begin{thebibliography}{10}

\bibitem{Bell1986QFTWithoutObservers}
J.S. Bell.
\newblock Quantum field theory without observers.
\newblock {\em Phys. Rep.}, 137(1):49--54, 1986.

\bibitem{Bohm1952SuggestedInterpretationOfQuantumMechanicsInTermsOfHiddenVariables}
D.~Bohm.
\newblock {A} suggested interpretation of quantum mechanics in terms of
  ``hidden'' variables, {I \& II}.
\newblock {\em Phys. Rev.}, 85(2):166--193, 1952.

\bibitem{Bohr1928ComoLecture}
N~Bohr.
\newblock The {Q}uantum {P}ostulate and the recent development of atomic
  theory.
\newblock {\em Supplement to "Nature"}, 121:580--590, 1928.
\newblock Reprinted in {Wheeler and Zurek, 1983}.

\bibitem{Bohr1935CanQuantumMechanicalDescriptionOfPhysicalRealityBeConsideredComplete}
Niels Bohr.
\newblock Can quantum-mechanical description of physical reality be considered
  complete?
\newblock {\em Phys. Rev.}, 48(8):696, 1935.

\bibitem{Carroll2021ConsciousnessAndTheLawsOfPhysics}
Sean Carroll.
\newblock Consciousness and the laws of physics.
\newblock {\em Journal of Consciousness Studies}, 28(9-10):16--31, 2021.

\bibitem{Carroll2021RealityAsAVectorInHilbertSpace}
S.M. Carroll.
\newblock Reality as a vector in {H}ilbert space.
\newblock In Valia Allori, editor, {\em Quantum mechanics and fundamentality},
  volume 460, pages 211--224. Springer Nature, 2022.

\bibitem{CarrollSingh2019MadDogEverettianism}
S.M. Carroll and A.~Singh.
\newblock Mad-dog {E}verettianism: {Q}uantum {M}echanics at its most minimal.
\newblock In A.~Aguirre, B.~Foster, and Z.~Merali, editors, {\em What is
  Fundamental?}, pages 95--104. Springer, 2019.

\bibitem{CarrollSingh2021QuantumMereology}
S.M. Carroll and A.~Singh.
\newblock Quantum mereology: {F}actorizing {H}ilbert space into subsystems with
  quasi-classical dynamics.
\newblock {\em To appear in Phys. Rev. A}, 2021.
\newblock Preprint \href{http://arxiv.org/abs/2005.12938}{arXiv:2005.12938}.

\bibitem{ChangAndKeisler1990ModelTheory}
C.C. Chang and H.J. Keisler.
\newblock {\em Model theory}.
\newblock Elsevier, 1990.

\bibitem{CotlerEtAl2019LocalityFromSpectrum}
J.S. Cotler, G.R. Penington, and D.H. Ranard.
\newblock Locality from the spectrum.
\newblock {\em Comm. Math. Phys.}, 368(3):1267--1296, 2019.

\bibitem{Dewitt1967QuantumTheoryOfGravityI_TheCanonicalTheory}
B.S. DeWitt.
\newblock Quantum theory of gravity. {I}. {T}he canonical theory.
\newblock {\em Phys. Rev.}, 160(5):1113, 1967.

\bibitem{Eddington1921SpaceTimeAndGravitation}
A.S. Eddington.
\newblock {\em Space, time and gravitation: {A}n outline of the general
  relativity theory}.
\newblock The University Press, 1921.

\bibitem{Eddington1934NewPathwaysInScience_Boltzmann_brains}
A.S. Eddington.
\newblock {\em New pathways in science}.
\newblock Cambridge Univ. Press, Cambridge, UK, 1934.

\bibitem{Elze2024CellularAutomatonOntologyBitsQubitsAndTheDiracEquation}
Hans-Thomas Elze.
\newblock Cellular automaton ontology, bits, qubits and the {D}irac equation.
\newblock {\em Int. J. Quantum Inf.}, 22(06):2450013, 2024.

\bibitem{Everett1957RelativeStateFormulationOfQuantumMechanics}
H.~Everett.
\newblock ``{R}elative state'' formulation of quantum mechanics.
\newblock {\em Rev. Mod. Phys.}, 29(3):454--462, Jul 1957.

\bibitem{GhirardiRiminiWeber1980AGeneralArgumentAgainstSuperluminalTransmissionThroughQuantumMechanicalMeasurementProcess}
G-C Ghirardi, A~Rimini, and T~Weber.
\newblock A general argument against superluminal transmission through the
  quantum mechanical measurement process.
\newblock {\em Lettere al Nuovo Cimento (1971-1985)}, 27(10):293--298, 1980.

\bibitem{Goldstein1998QuantumTheoryWithoutObserversI}
S.~Goldstein.
\newblock Quantum theory without observers.
\newblock {\em Physics Today}, 51(3):42--47, 1998.

\bibitem{Goldstein1998QuantumTheoryWithoutObserversII}
S.~Goldstein.
\newblock Quantum theory without observers, {P}art two.
\newblock {\em Physics Today}, 51(4):38--42, 1998.

\bibitem{Gratzer2008UniversalAlgebra}
G.~Gr{\"a}tzer.
\newblock {\em Universal algebra}.
\newblock Springer Science \& Business Media, 2008.

\bibitem{Hatfield2018QuantumFieldTheoryOfPointParticlesAndStrings}
Brian Hatfield.
\newblock {\em Quantum field theory of point particles and strings}.
\newblock CRC Press, Boca Raton, FL, USA, 2018.

\bibitem{Heisenberg1958PhysicsAndPhilosophy}
W.~Heisenberg.
\newblock {\em Physics and Philosophy: {T}he Revolution in Modern Science}.
\newblock Harper \& Brothers Publishers, New York, 1958.

\bibitem{sep-structural-realism}
J.~Ladyman.
\newblock Structural realism.
\newblock In Edward~N. Zalta, editor, {\em The Stanford Encyclopedia of
  Philosophy}. Metaphysics Research Lab, Stanford University, Stanford,
  California, spring 2020 edition, 2020.

\bibitem{sep-functionalism}
J.~Levin.
\newblock Functionalism.
\newblock In E.N. Zalta, editor, {\em The {Stanford} Encyclopedia of
  Philosophy}. Metaphysics Research Lab, Stanford University, Stanford,
  California, fall 2018 edition, 2018.

\bibitem{Schlosshauer2007DecoherenceAndTheQuantumToClassicalTransition}
M.~M.~Schlosshauer.
\newblock {\em Decoherence and the Quantum-To-Classical Transition}.
\newblock Springer, Berlin, 2007.

\bibitem{Maldacena1999AdSCFTInitial}
J~Maldacena.
\newblock The large-{N} limit of superconformal field theories and
  supergravity.
\newblock {\em Int. J. Theor. Phys}, 38(4):1113--1133, 1999.

\bibitem{Piazza2010GlimmersOfAPreGeometricPerspective}
F.~Piazza.
\newblock Glimmers of a pre-geometric perspective.
\newblock {\em Found. Phys.}, 40:239--266, 2010.

\bibitem{Piccinini2004FunctionalismComputationalismAndMentalStates}
G.~Piccinini.
\newblock Functionalism, computationalism, and mental states.
\newblock {\em Stud. Hist. Philos. Sci. A}, 35(4):811--833, 2004.

\bibitem{sep-materialism-eliminative}
William Ramsey.
\newblock {Eliminative Materialism}.
\newblock In Edward~N. Zalta, editor, {\em The {Stanford} Encyclopedia of
  Philosophy}. Metaphysics Research Lab, Stanford University, Stanford,
  California, {S}pring 2022 edition, 2022.

\bibitem{sep-computational-mind}
M.~Rescorla.
\newblock {The Computational Theory of Mind}.
\newblock In Edward~N. Zalta, editor, {\em The {Stanford} Encyclopedia of
  Philosophy}. Metaphysics Research Lab, Stanford University, Stanford,
  California, fall 2020 edition, 2020.

\bibitem{Schulman1984DefiniteMeasurementsAndDeterministicQuantumEvolution}
L.S. Schulman.
\newblock Definite measurements and deterministic quantum evolution.
\newblock {\em Phys. Lett. A}, 102(9):396--400, 1984.

\bibitem{Schulman2017ProgramSpecialState}
L.S. Schulman.
\newblock Program for the special state theory of quantum measurement.
\newblock {\em Entropy}, 19(7):343, 2017.

\bibitem{Stoica2008FlowingWithAFrozenRiver}
O.C. Stoica.
\newblock Flowing with a frozen river. {E}ssay awarded with a mention.
\newblock {\em Foundational Questions Institute,
  \href{http://fqxi.org/community/essay/winners/2008.1}{``The Nature of Time''
  essay contest}}, 2008.
\newblock
  \href{http://fqxi.org/community/forum/topic/322}{http://fqxi.org/community/forum/topic/322},
  last accessed \today.

\bibitem{Stoica2008SmoothQuantumMechanics}
O.C. Stoica.
\newblock Smooth quantum mechanics.
\newblock {\em PhilSci Archive}, 2008.
\newblock
  \href{http://philsci-archive.pitt.edu/archive/00004344/}{philsci-archive:00004344}.

\bibitem{Stoica2012QMGlobalAndLocalAspectsOfCausalityInQuantumMechanics}
O.C. Stoica.
\newblock \href{http://dx.doi.org/10.1051/epjconf/20135801017}{Global and local
  aspects of causality in quantum mechanics}.
\newblock In {\em
  \href{http://www.epj-conferences.org/index.php?option=com_toc&url=/articles/epjconf/abs/2013/19/contents/contents.html}{EPJ
  Web of Conferences}, TM 2012 -- The Time Machine Factory [unspeakable,
  speakable] on Time Travel in Turin}, volume~58, page 01017, 9 2013.

\bibitem{Stoica2012QMQuantumMeasurementAndInitialConditions}
O.C. Stoica.
\newblock Quantum measurement and initial conditions.
\newblock {\em Int. J. Theor. Phys.}, pages 1--15, 2015.
\newblock \href{http://arxiv.org/abs/1212.2601}{arXiv:quant-ph/1212.2601}.

\bibitem{Stoica2016OnTheWavefunctionCollapse}
O.C. Stoica.
\newblock On the wavefunction collapse.
\newblock {\em Quanta}, 5(1):19--33, 2016.
\newblock
  \href{http://dx.doi.org/10.12743/quanta.v5i1.40}{http://dx.doi.org/10.12743/quanta.v5i1.40}.

\bibitem{Stoica2017TheUniverseRemembersNoWavefunctionCollapse}
O.C. Stoica.
\newblock The universe remembers no wavefunction collapse.
\newblock {\em Quantum Stud. Math. Found.}, 2017.
\newblock \href{http://arxiv.org/abs/1607.02076}{arXiv:1607.02076}.

\bibitem{Stoica2021PostDeterminedBlockUniverse}
O.C. Stoica.
\newblock The post-determined block universe.
\newblock {\em Quantum Stud. Math. Found.}, 8(1), 2021.
\newblock \href{http://arxiv.org/abs/1903.07078}{arXiv:1903.07078}.

\bibitem{Stoica2020StandardQuantumMechanicsWithoutObservers}
O.C. Stoica.
\newblock Standard quantum mechanics without observers.
\newblock {\em Phys. Rev. A}, 103(3):032219, 2021.

\bibitem{Stoica2022SpaceThePreferredBasisCannotUniquelyEmergeFromTheQuantumStructure}
O.C. Stoica.
\newblock 3d-space and the preferred basis cannot uniquely emerge from the
  quantum structure.
\newblock {\em Adv. Theor. Math. Phys.}, 26(10):3895---3962, 2022.
\newblock \href{http://arxiv.org/abs/2102.08620}{arXiv:2102.08620}.

\bibitem{Stoica2023AreObserversReducibleToStructures}
O.C. Stoica.
\newblock Are observers reducible to structures?
\newblock 2023.

\bibitem{Stoica2023NoGoResultsOnEmergentSpaceAndOtherStructures}
O.C. Stoica.
\newblock No-go results on emergent space and other structures.
\newblock {\em Journal of Physics: Conference Series}, 2533(1):012027, 2023.

\bibitem{Stoica2023PrinceAndPauperQuantumParadoxHilbertSpaceFundamentalism}
O.C. Stoica.
\newblock The prince and the pauper. {A} quantum paradox of {H}ilbert-space
  fundamentalism.
\newblock {\em Preprint
  \href{https://arxiv.org/abs/2310.15090}{arXiv:2310.15090}}, 2023.

\bibitem{Stoica2024DoesTheHamiltonianDetermineTheTPSAndThe3dSpace}
O.C. Stoica.
\newblock Does the {H}amiltonian determine the tensor product structure and the
  3d space?
\newblock {\em Preprint
  \href{https://arxiv.org/abs/2401.01793}{arXiv:2401.01793}}, 2024.

\bibitem{Stoica2024EmpiricalAdequacyOfTimeOperatorCC2HamiltonianGeneratingTranslations}
O.C. Stoica.
\newblock Empirical adequacy of the time operator canonically conjugate to a
  hamiltonian generating translations.
\newblock {\em Physica Scripta}, 2024.
\newblock \href{https://arxiv.org/abs/2204.01426}{arXiv:2204.01426}.

\bibitem{Stoica2024FreedomInTheManyWorldsInterpretation}
O.C. Stoica.
\newblock Freedom in the many-worlds interpretation.
\newblock {\em Found. Phys.}, 54(5):68, 2024.

\bibitem{Stoica2024ObservationAsPhysication}
O.C. Stoica.
\newblock Observation as physication. {A} single-world unitary no-conspiracy
  interpretation of quantum mechanics.
\newblock {\em Preprint
  \href{https://arxiv.org/abs/2412.09669}{arXiv:2412.09669}}, 2024.

\bibitem{sep-physicalism}
Daniel Stoljar.
\newblock {Physicalism}.
\newblock In Edward~N. Zalta and Uri Nodelman, editors, {\em The {Stanford}
  Encyclopedia of Philosophy}. Metaphysics Research Lab, Stanford University,
  Stanford, California, {S}ummer 2023 edition, 2023.

\bibitem{Strawson2006RealisticMonismWhyPhysicalismEntailsPanpsychism}
Galen Strawson.
\newblock Realistic monism: {W}hy physicalism entails panpsychism.
\newblock In A.~Freeman, editor, {\em Consciousness and its place in nature:
  {D}oes physicalism entail panpsychism?}, pages 3--31, UK, 2006. Imprint
  Academic.

\bibitem{tHooft2016CellularAutomatonInterpretationQM}
G.~'t~Hooft.
\newblock {\em The cellular automaton interpretation of quantum mechanics},
  volume 185.
\newblock Springer, 2016.

\bibitem{vonNeumann1955MathFoundationsQM}
J.~{von Neumann}.
\newblock {\em Mathematical Foundations of Quantum Mechanics}.
\newblock Princeton University Press, Princeton, NJ, 1955.

\bibitem{Wallace2022StatingStructuralRealismMathematicsFirstApproachesToPhysicsAndMetaphysics}
D.~Wallace.
\newblock Stating structural realism: mathematics-first approaches to physics
  and metaphysics.
\newblock {\em Philosophical Perspectives}, 36(1):345--378, 2022.

\bibitem{Wigner1967RemarksOnTheMindBodyProblemWignersFriend}
E.P. Wigner.
\newblock Remarks on the mind-body problem.
\newblock In {\em Symmetries and reflections: {S}cientific essays}, pages
  171--184. Indiana University Press, Bloomington, 1967.

\bibitem{Worrall1989StructuralRealism}
J.~Worrall.
\newblock Structural realism: {T}he best of both worlds?
\newblock {\em Dialectica}, 43(1-2):99--124, 1989.

\bibitem{ZanardiLidarLloyd2004QuantumTensorProductStructuresAreObservableInduced}
P.~Zanardi, D.A. Lidar, and S.~Lloyd.
\newblock Quantum tensor product structures are observable induced.
\newblock {\em Phys. Rev. Lett.}, 92(6):060402, 2004.

\bibitem{Zurek2003DecoherenceEinselectionAndTheQuantumOriginsOfTheClassical}
W.~H. Zurek.
\newblock {Decoherence, Einselection, and the Quantum Origins of the
  Classical}.
\newblock {\em Rev.\ Mod.\ Phys.}, 75:715, 2002.
\newblock \href{http://arxiv.org/abs/quant-ph/0105127}{arXiv:quant-ph/0105127}.

\bibitem{Zurek1991DecoherenceAndTheTransitionFromQuantumToClassical}
W.H. Zurek.
\newblock Decoherence and the transition from quantum to classical.
\newblock {\em Physics Today}, page~37, 1991.

\end{thebibliography}
\end{document}